\documentclass[aps,prb,superscriptaddress,twocolumn]{revtex4}
\usepackage{amsmath,amsfonts,bm,graphicx,color}

\begin{document}

\title{Measurement of finite-frequency current statistics in a single-electron transistor}
\author{Niels Ubbelohde}
\affiliation{Institut f\"ur Festk\"orperphysik, Leibniz Universit\"at Hannover, 30167 Hannover, Germany}
\author{Christian Fricke}
\affiliation{Institut f\"ur Festk\"orperphysik, Leibniz Universit\"at Hannover, 30167 Hannover, Germany}
\author{Christian Flindt}
\affiliation{D\'epartement de Physique Th\'eorique, Universit\'e de Gen\`eve, 1211 Gen\`eve, Switzerland}
\affiliation{Department of Physics, Harvard University, Cambridge, MA 02138, USA}
\author{Frank Hohls}
\affiliation{Institut f\"ur Festk\"orperphysik, Leibniz Universit\"at Hannover, 30167 Hannover, Germany}
\affiliation{Physikalisch-Technische Bundesanstalt, 38116 Braunschweig, Germany}
\author{Rolf J.\ Haug}
\affiliation{Institut f\"ur Festk\"orperphysik, Leibniz Universit\"at Hannover, 30167 Hannover, Germany}

\newcommand{\llangle}{\langle\!\langle}
\newcommand{\rrangle}{\rangle\!\rangle}

\date{\today}
\maketitle

{\bf
Electron transport in nano-scale structures is strongly influenced by the Coulomb interaction which gives rise to correlations in the stream of charges and leaves clear fingerprints in the fluctuations of the electrical current. A complete understanding of the underlying physical processes requires measurements of the electrical fluctuations on all time and frequency scales, but experiments  have so far been restricted to fixed frequency ranges as broadband detection of current fluctuations is an inherently difficult experimental procedure. Here we demonstrate that the electrical fluctuations in a single electron transistor (SET) can be accurately measured on all relevant frequencies using a nearby quantum point contact for on-chip real-time detection of the current pulses in the SET. We have directly measured the frequency-dependent current statistics and hereby fully characterized the fundamental tunneling processes in the SET. Our experiment paves the way for future investigations of interaction and coherence induced correlation effects in quantum transport.
}
\newline


The electrical fluctuations in a nano-scale conductor reveal a wealth of information about the physical processes inside the device compared to what is available from a conductance measurement alone.\cite{Blanter2000,Nazarov2003,Levitov1993} Zero-frequency noise measurements are now routinely performed using standard techniques, but more recently the detection of higher-order current correlation functions, or cumulants, has attracted considerable attention in experimental\cite{Reulet2003,Bylander2005,Bomze2005,Gustavsson2006,Fujisawa2006,Sukhorukov2007,Fricke2007,Timofeev2007,Gershon2008,Gustavsson2009,Flindt2009,Gabelli2009a,Zhang2009,Gabelli2009b,Xue2009,Mahe2010} and theoretical\cite{Tobiska2004,Ankerhold2005,Zazunov2007} studies of charge transport in man-made sub-micron structures. Measurements have for example been encouraged by the intriguing connections between current fluctuations and entanglement entropy in solid-state systems\cite{Klich2009,Song2010} as well as by the possibility to test fluctuation theorems\cite{Foerster2008,Sanchez2010,Nakamura2010} at the nano-scale which are now a topic at the forefront of non-equilibrium statistical physics.

Experiments on current fluctuations have mainly focused on zero-frequency current correlations\cite{Reulet2003,Bylander2005,Bomze2005,Gustavsson2006,Fujisawa2006,Sukhorukov2007,Fricke2007,Timofeev2007,Gershon2008,Gustavsson2009,Flindt2009,Gabelli2009a,Zhang2009} while measurements of finite-frequency cumulants of the current have remained an outstanding experimental challenge. However, noise detection in a restricted frequency band gives only partial information about correlations and measurements that cover the full frequency range are necessary to access all relevant time scales that characterize the transport process. In a zero-frequency measurement, correlation effects are integrated over a long period of time and important information about characteristic time scales are lost. To observe the \emph{dynamical} features of the correlations, finite-frequency measurements are required.\cite{Levitov1993,Nazarov2003,Galaktionov2003,Nagaev2004,Pilgram2004,Salo2006,Emary2007,Marcos2010,Galaktionov2011}
It has even been shown theoretically that the frequency-dependent third cumulant (the skewness) of the current may contain additional information about correlations and internal time scales of a system compared to the finite-frequency noise alone, for instance in chaotic cavities\cite{Nagaev2004} and diffusive conductors.\cite{Pilgram2004}

\begin{figure*}
\begin{center}
\includegraphics[width=0.98\linewidth]{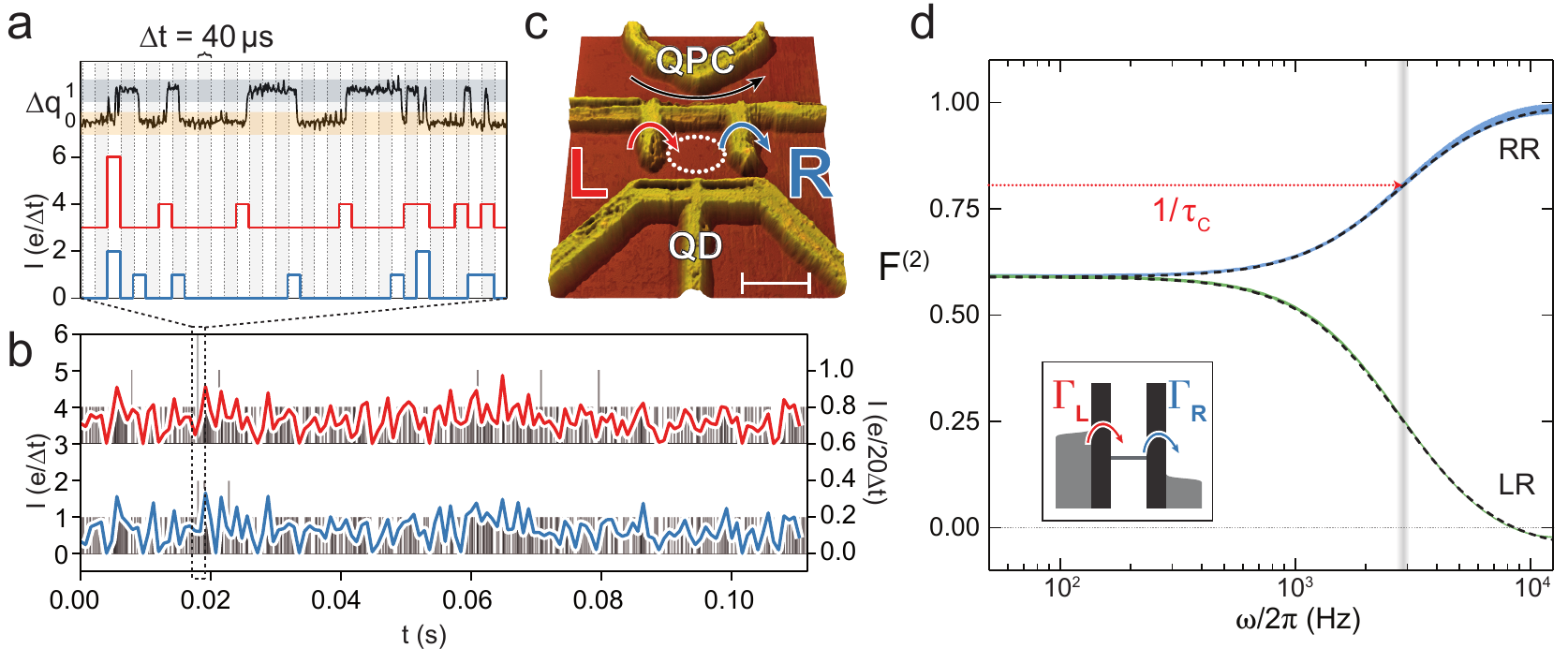}
\caption{{\bf Single electron transistor (SET) and finite-frequency noise spectra.} {\bf a,b}, Fluctuations of the current through the left (red curve) and right (blue curve) tunneling barriers of the SET. {\bf c}, Atomic force microscope topography of the SET consisting of a Coulomb blockade quantum dot (QD) coupled via tunneling barriers to left ($L$) and right ($R$) electrodes. The scale bar corresponds to 250 nm. A bias difference between the electrodes drives a stream of electrons through the QD from $L$ to $R$. An electron entering the QD from $L$ causes a suppression of the current through the nearby quantum point contact (QPC) until the electron tunnels out of the QD to $R$. The occupation of the QD ($\Delta q=0,1$) is inferred from the time-dependent current through the QPC (shown with black in {\bf a}). The corresponding pulse currents through the left (red) and right (blue) barriers are discretized in time steps of $\Delta t=40$ $\mu$s
and shown displaced for clarity, {\bf a}. In {\bf b}, the same currents are shown on a much longer time scale and discretized in time steps of $\Delta t$ (grey spikes) and $20\Delta t$ (full lines), respectively.  {\bf d}, Noise spectrum (blue curve) of the current through the right barrier ($RR$) and cross-correlations (green curve) between the two currents ($LR$). The thickness of the lines indicate experimental error estimates. Model calculations (dashed lines) of the spectra are in excellent agreement with the experiment using $\Gamma_L=13.23$ kHz and $\Gamma_R=4.81$ kHz for the tunneling rates across the barriers and a detector rate of $\Gamma_D=0.3$ MHz. The mean dwell time of electrons on the QD is $\tau_d=\Gamma_R^{-1}\simeq  210$ $\mu$s and the correlation time is $\tau_c=(\Gamma_L+\Gamma_R)^{-1}\simeq 55$ $\mu$s. The inset shows a schematic model of the charge transport process.} \label{Fig1}
\end{center}
\end{figure*}

Figure \ref{Fig1}a-c shows a typical time trace of the currents in our nano-scale SET consisting of a quantum dot (QD) coupled via tunneling barriers to source and drain electrodes. The QD is operated close to a charge degeneracy point in the Coulomb blockade regime, where only a single additional electron at a time can enter from the source electrode and leave via the drain electrode. The applied voltage bias $eV$ is larger than the electronic temperature $k_BT$ which prevents electrons from tunneling in the opposite direction of the mean current. While a stream of electrons is driven through the QD, a separate current is passed through the nearby quantum point contact (QPC) whose conductance is highly sensitive to the presence of additional electrons ($\Delta q=0,1$) on the QD.\cite{Lu2003,Vandersypen2004} By monitoring the switches of the current through the QPC we can thus detect in real-time single electrons tunneling through the left ($L$) and right ($R$) barriers and thereby determine the corresponding pulse currents $I_L(t)$ and $I_R(t)$. The time trace of the currents illustrates the electrical fluctuations of interest here.

We now analyze the current fluctuations obtained from data measured over 24 hours during which 300 million electrons passed through the SET. Our high-quality measurements allow us to develop a complete picture of the noise properties of the SET, which does not only focus on the zero-frequency components of the fluctuations, but contains the full frequency-resolved information about the noise and higher-order correlation functions. From the measured finite-frequency noise spectrum we extract the correlation time of the current fluctuations. The noise spectrum, however, is a one-frequency quantity only which does not reflect correlations between different spectral components of the current. To observe such correlations, we employ bispectral analysis and consider the finite-frequency skewness (or bispectrum) of the current. The skewness shows that the current fluctuations are non-gaussian on all relevant time and frequency scales due to the non-equilibrium conditions imposed by the applied voltage bias. Our measurements are supported by model calculations which are in excellent agreement with the experimental data. The results presented here provide a fundamental understanding of the electrical fluctuations in SETs which are expected to constitute the basic building blocks of future nano-scale electronics.
\newline

\noindent {\bf Results}
\newline

\begin{figure*}
\begin{center}
\includegraphics[width=0.98\linewidth]{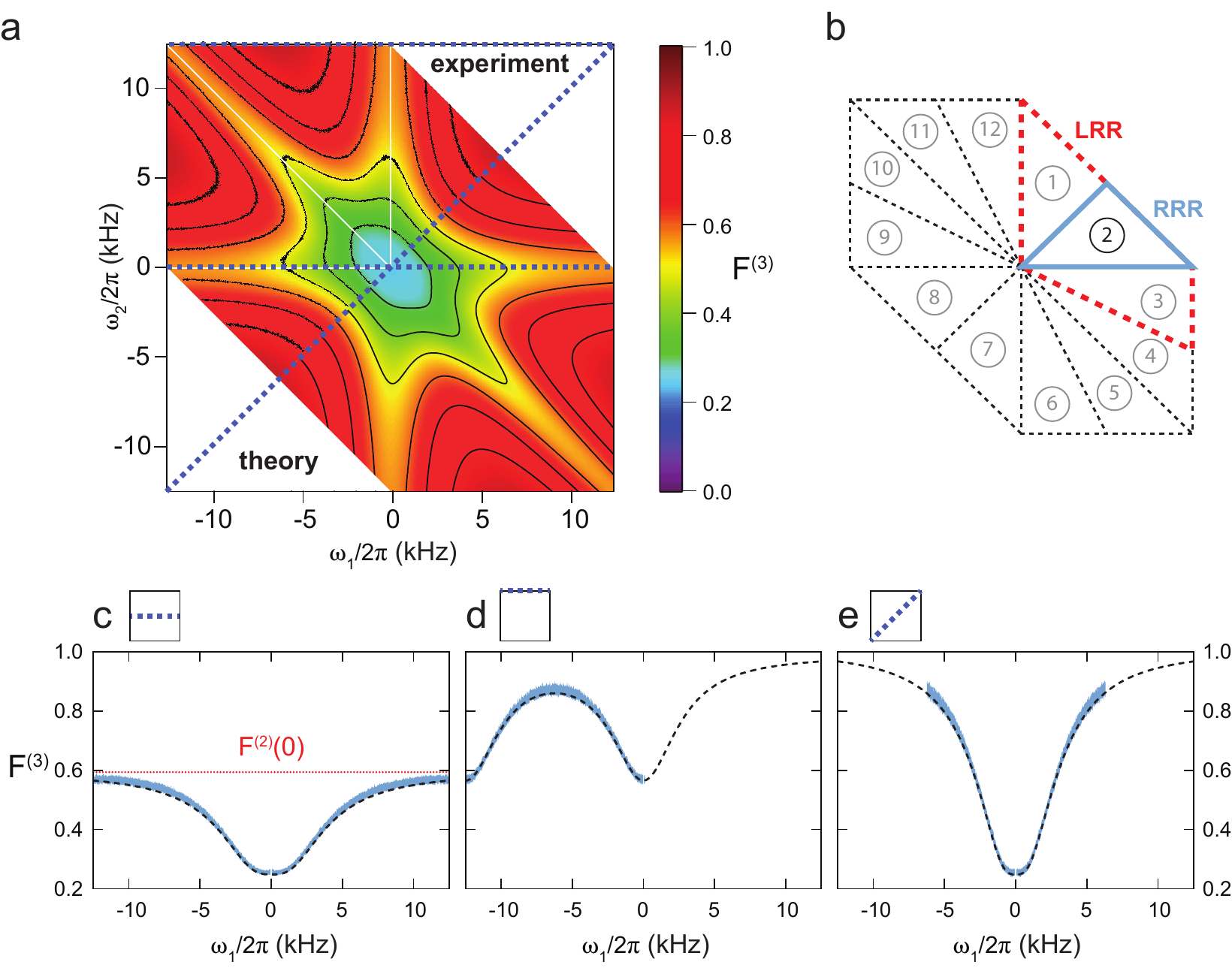}
\caption{{\bf Finite-frequency skewness.} {\bf a}, Experimental results and model calculations of the third Fano factor (the frequency-dependent skewness) $F^{(3)}(\omega_1,\omega_{2})=S^{(3)}(\omega_1,\omega_2)/e^2I$. Experimental results are shown above the diagonal $\omega_1=\omega_2$ and model calculations below. Contour lines indicate $F^{(3)}=0.3$, $0.4$, $0.5$, $0.6$, $0.7$, and $0.8$. The parameters for the model calculations are given in the caption of Fig.\ \ref{Fig1}. {\bf b}, Symmetries of the bispectrum $S^{(3)}(\omega_1,\omega_2)$. Several important symmetry conditions follow from the definition,\cite{Nikias1993} including symmetry with respect to interchange of frequencies $S^{(3)}(\omega_1,\omega_2)=S^{(3)}(\omega_2,\omega_1)$. Knowledge of the bispectrum in any of the regions $\textcircled{\footnotesize{1}}-\textcircled{\footnotesize{12}}$ is sufficient for a complete description of the bispectrum. We have measured the skewness in region $\textcircled{\footnotesize{2}}$ and the cross-bispectrum (Fig.\ \ref{Fig3}) in regions $\textcircled{\footnotesize{1}}-\textcircled{\footnotesize{3}}$. {\bf c,d,e}, Finite-frequency skewness along the three blue lines  in {\bf a}. Full lines are experimental results, while dashed lines show model calculations. The thickness of the full lines indicate the experimental error estimates. The third order Fano factor $F^{(3)}(\omega_1,\omega_{2})$ approaches the zero-frequency shot noise $F^{(2)}(0)$ for $\omega_2=0$ and $|\omega_1|$ being much larger than the inverse correlation time $\tau_c^{-1}$ as seen in {\bf c}.} \label{Fig2}
\end{center}
\end{figure*}

{\bf Noise spectrum.} The measured noise spectrum for the right tunneling barrier is presented in Fig.\ \ref{Fig1}d. The finite-frequency current-correlation function $S_{\alpha\beta}^{(2)}(\omega)$ is defined in terms of the noise power as\cite{Blanter2000}
\begin{equation}
\llangle \hat{I}_\alpha(\omega) \hat{I}_\beta(\omega')\rrangle= 2\pi \delta(\omega+\omega')S_{\alpha\beta}^{(2)}(\omega),
\nonumber
\end{equation}
where $\hat{I}_\alpha(\omega)$, $\alpha=L,R$, is the Fourier transformed current and double brackets $\langle\!\langle\ldots\rangle\!\rangle$ denote cumulant averaging over many experimental realizations. Figure \ref{Fig1}d shows the Fano factor $F^{(2)}(\omega)=S_{RR}^{(2)}(\omega)/eI$ for the right tunneling barrier with the mean current $I=\langle I_L(t)\rangle=\langle I_R(t)\rangle$ being constant in the stationary state.  The noise is symmetric in frequency, $S_{RR}^{(2)}(\omega)=S_{RR}^{(2)}(-\omega)$, and results are shown for positive frequencies only. For uncorrelated transport  the noise spectrum would be white, i.\ e.,  $F^{(2)}(\omega)=1$ on all frequencies, corresponding to a Poisson process. Our measurements, in contrast, show a clear suppression of fluctuations below the Poisson value at low frequencies. This is due to the strong Coulomb interactions on the QD which introduce correlations in the stream of electrons propagating through the SET: each electron spends a finite time on the QD during which it blocks the next electron entering the QD. The measured noise spectrum has a Lorentzian shape whose width is determined by the inverse dynamical time scale of the correlations. Our measurements of the finite-frequency noise thereby enable a \emph{direct} observation of the correlation time $\tau_c\simeq 55$ $\mu s$ of the transport, Fig.~\ref{Fig1}d.

To corroborate our experimental results we calculate the noise spectrum of the schematic model in the inset of Fig.\ \ref{Fig1}d. Single electrons tunnel from the left electrode onto the QD at rate $\Gamma_L$ and leave it via the right electrode at rate $\Gamma_R$. The Fano factor is then\cite{Blanter2000}
\begin{equation}
F^{(2)}(\omega)=1-\frac{2\Gamma_L\Gamma_R}{(\Gamma_L+\Gamma_R)^2+\omega^2}
\nonumber
\end{equation}
where theoretically $\tau_c=(\Gamma_L+\Gamma_R)^{-1}$ is identified as the correlation time. This expression qualitatively explains the measured noise spectrum. Quantitative agreement is obtained by also taking into account the finite detection rate $\Gamma_D$ of the QPC charge sensing scheme.\cite{Naaman2006,Gustavsson2009} The three parameters $\Gamma_L$, $\Gamma_R$, and $\Gamma_D$ can be independently extracted from the distribution of waiting times between detected tunneling events.\cite{Brandes2008,Flindt2009,Albert2011} The model calculations (see {\bf Methods}) are in excellent agreement with measurements over the full range of frequencies, demonstrating the high quality of our experimental data.

\begin{figure*}
\begin{center}
\includegraphics[width=0.95\linewidth, trim = 0 0 0 0, clip]{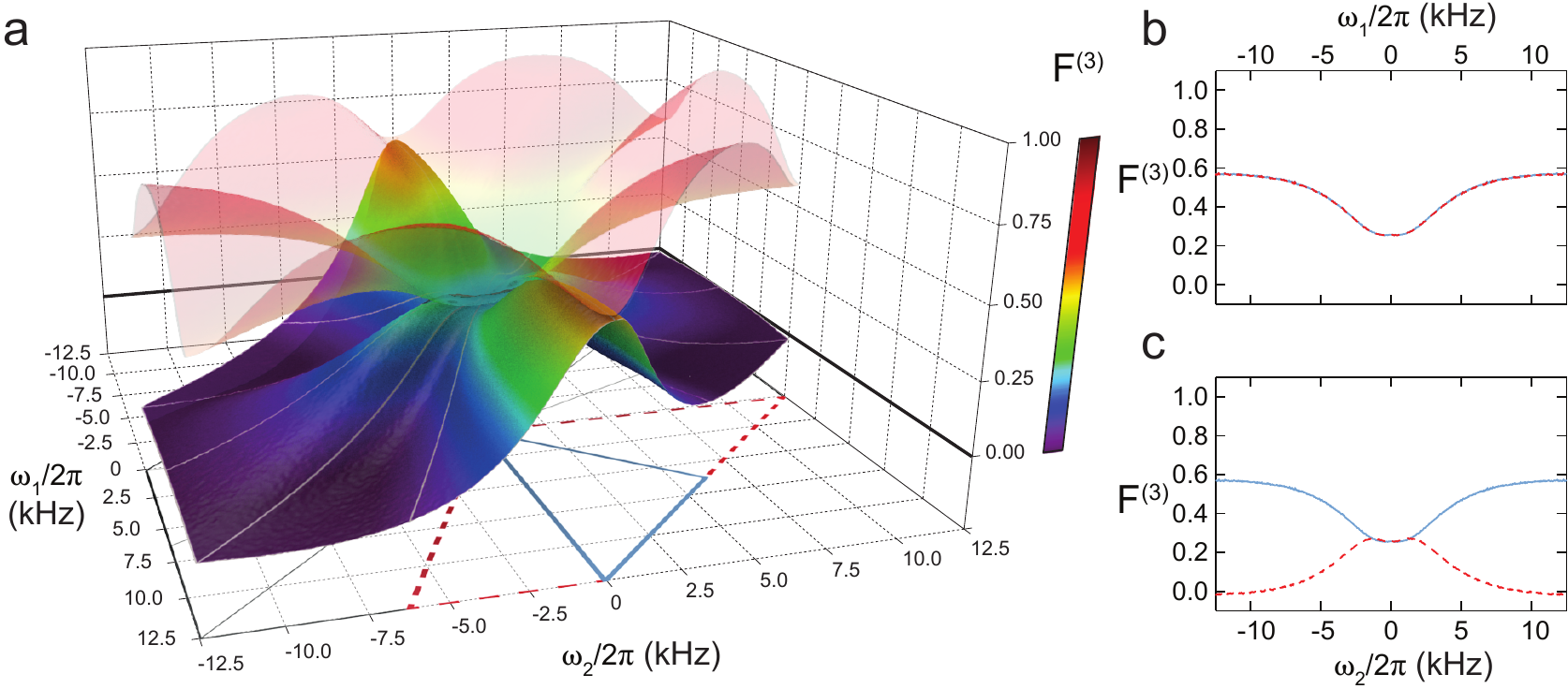}
\caption{{\bf Finite-frequency cross-bispectrum.} {\bf a,} Measurements of the frequency-dependent cross-bispectrum $F^{(3)}_{LRR}(\omega_1,\omega_2)$. The cross-bispectrum shows a reduced symmetry compared to the bispectrum and measurements in the regions $\textcircled{\footnotesize{1}}$-$\textcircled{\footnotesize{3}}$, Fig.\ \ref{Fig2}b, are required for a complete characterization of the cross-bispectrum. For comparison, the bispectrum $F^{(3)}(\omega_1,\omega_{2})$ from Fig.\ \ref{Fig2}a is shown as a semi-transparent surface above the cross-bispectrum. {\bf b,} Bispectrum and cross-bispectrum along the line $\omega_2=0$, where they coincide. {\bf c,} Bispectrum (full blue line) and cross-bispectrum (dashed red line) along the line $\omega_1=0$.} \label{Fig3}
\end{center}
\end{figure*}

{\bf Cross-correlations.} Cross-correlations between the left $I_L(t)$ and the right $I_R(t)$ currents can also be measured, Fig.\ \ref{Fig1}d. For the schematic model, the (cross-correlation) Fano factor reads $F^{(2)}_{LR}(\omega)\equiv \mathrm{Re}[S^{(2)}_{LR}(\omega)]/eI=(\Gamma_L^2+\Gamma_R^2)/([\Gamma_L+\Gamma_R]^2+\omega^2)$, which in the zero-frequency limit coincides with the noise spectrum, $F^{(2)}_{LR}(0)=F^{(2)}(0)$, as a consequence of charge conservation on the QD. The two currents are clearly correlated at frequencies that are lower than the inverse correlation time $\tau_c^{-1}$, but the cross-correlator eventually reaches zero at higher frequencies. Interestingly, the cross-correlations of the detected pulse currents are slightly negative at high frequencies. This is due to the finite resolution of the QPC charge sensing protocol which is not able to distinguish current pulses that are separated in time by an interval which is shorter than the inverse detector rate $\Gamma_D^{-1}$.

{\bf Higher-order cumulants.} We now turn to measurements of higher-order finite-frequency cumulants. The $m$'th finite-frequency current correlator corresponding to a time-dependent current $I(t)$ is defined as
\begin{equation}
\llangle \hat{I}(\omega_1)\cdots \hat{I}(\omega_{m})\rrangle=2\pi \delta(\omega_1+\ldots+\omega_m) S^{(m)}(\omega_1,\ldots,\omega_{m-1})
\nonumber
\end{equation}
where translational invariance in time implies frequency conservation as indicated by the Dirac delta function $\delta(\omega)$ and $S^{(m)}(\omega_1,\ldots,\omega_{m-1})$ is the polyspectrum.\cite{Nikias1993,Kogan1996} In the case $m=2$, the polyspectrum yields the noise power spectrum $S^{(2)}(\omega)$, while the skewness (or bispectrum) is given by $m=3$ with the corresponding Fano factor $F^{(3)}(\omega_1,\omega_2)= S^{(3)}(\omega_1,\omega_{2})/e^{2}I$.  We focus here on the frequency-dependent skewness $S^{(3)}(\omega_1,\omega_2)$, although our experimental data in principle allows us also to extract cumulants of even higher orders.

The measured skewness, Fig.\ \ref{Fig2}a, shows a much richer structure and frequency dependence compared to the noise spectrum. The skewness obeys several symmetries following from the definition, Fig.\ \ref{Fig2}b. We exploit the mirror symmetry with respect to interchange of frequencies, $S^{(3)}(\omega_1,\omega_2)=S^{(3)}(\omega_2,\omega_1)$, to compare measurement and model calculations: experimental results are presented above the diagonal $\omega_1=\omega_2$, while model calculations are shown below, Fig.\ \ref{Fig2}a. The nonzero bispectrum indicates non-Gaussian statistics on all frequencies and shows strong correlations between different spectral components of the current. Importantly, these correlations are not a consequence of non-linearities in the detection scheme, but are solely due to the physical non-equilibrium conditions imposed by the applied voltage bias. The pulse currents are directly derived from the tunneling events and the influence of external noise sources, including the amplification of the QPC current, is thereby explicitly avoided. Intuitively, one would expect the correlations to vanish, if the observation frequency is larger than the average frequency of the transport. Surprisingly, however, a certain degree of correlation persists even if one frequency is large, while the other is kept finite.  This is in stark contrast to the second Fano factor $F^{(2)}(\omega)$ which approaches unity in the high-frequency limit, Fig. 1d, corresponding to uncorrelated tunneling events.

For the schematic model in Fig.\ \ref{Fig1}d, the finite-frequency skewness reads\cite{Emary2007}
\begin{equation}
F^{(3)}(\omega_1,\omega_2)=1-2\Gamma_L\Gamma_R\frac{\prod_{j=1}^2(\gamma_j^{2}+\omega_3^2-\omega_1\omega_2)}{\prod_{i=1}^3[(\Gamma_L+\Gamma_R)^{2}+\omega_i^2]},
\nonumber
\end{equation}
having defined $\gamma_1^2=\Gamma_L^2+\Gamma_R^2$, $\gamma_2^2=3(\Gamma_L+\Gamma_R)^2$, and $\omega_3=\omega_1+\omega_2$. This expression qualitatively explains the measured finite-frequency skewness and shows that the skewness has a complex structure which is not just a simple Lorentzian-shaped function of frequencies.  We note that the third Fano factor $F^{(3)}(\omega_1,\omega_2)$ reduces to the zero-frequency limit of the noise $F^{(2)}(0)$ in the limit $\omega_2 = 0$ and $|\omega_1|\gg \tau_c^{-1}$. This is immediately visible in Fig.\ \ref{Fig2}c. Again, quantitative agreement is achieved by including the finite detection rate $\Gamma_D$ in the model calculations as illustrated explicitly in Figs.\ \ref{Fig2}c-e. The analytic expression shows that the frequency dependence of the skewness, unlike the auto- and cross-correlation noise spectra, is not only governed by the correlation time $\tau_c=(\Gamma_L+\Gamma_R)^{-1}$. The skewness has an involved frequency dependence, given by several frequency scales, which cannot be deduced from the noise spectrum alone.

{\bf Higher-order cross-correlations.} In Fig.\ \ref{Fig3}a, we finally consider the cross-bispectrum $F^{(3)}_{LRR}(\omega_1,\omega_2)\equiv \mathrm{Re}[S^{(3)}_{LRR}(\omega_1,\omega_2)]/e^2I$,
measuring the third-order correlations between tunneling electrons entering and leaving the SET. The three non-redundant permutations of the left and right pulse currents lead to a reduced symmetry of the cross-bispectrum, as visualized in comparison with the auto-correlation bispectrum in Fig.\ \ref{Fig3}a. The cross-bispectrum coincides with the auto-bispectrum on the line $\omega_2 = 0$, where they approach the zero-frequency correlation of the shot noise for $|\omega_1|>\tau_c^{-1}$, Fig.\ \ref{Fig3}b. In contrast, for $|\omega_2|\neq 0$, the cross-bispectrum shows a frequency dependence similar to the cross-correlation shot noise and the anti-correlations due to the detection process become visible, Fig.\ \ref{Fig3}c.
\newline

\noindent {\bf Discussion}
\newline

We have measured the current statistics of charge transport in an SET and directly determined the dynamical features and time scales of the current fluctuations. From the measured frequency-dependent noise power we extracted the correlation time of the fluctuations. The noise power is a single-frequency quantity only and in order to investigate the correlations between different spectral components of the current using bispectral analysis we measured the frequency-dependent third order correlation function (the skewness). The skewness shows that the current statistics are non-gaussian on all frequencies due to the applied voltage bias. Our experimental results are supported by model calculations that are in excellent agreement with measurements. The results presented here are important for future applications of SETs in nano-scale electrical circuits operating with single electrons. Our accurate and stable experiment also facilitates several promising directions for basic research on nano-scale quantum devices. These include experimental investigations of fluctuation relations at finite frequencies and higher-order noise detection of interaction\cite{Barthold2006} and coherence\cite{Kiesslich2007} induced correlation effects in quantum transport under non-equilibrium conditions.
\newline

\noindent {\bf Methods}
\newline

\noindent {\bf Device fabrication.} The device was fabricated by local anodic oxidation techniques using an atomic force microscope on the surface of a GaAs/AlGaAs heterostructure with electron density $4.6\times 10^{15}$ m$^{-2}$ and a mobility of $64$ m$^2$/Vs. The two-dimensional electron gas residing 34 nm below the heterostructure surface is depleted underneath the oxidized lines, allowing us to define the quantum dot (QD) and the quantum point contact (QPC).

\noindent {\bf Measurements.} The experiment was carried out in a $^3$He cryostat at 500 mK with an applied bias of 900 $\mu$V across the QD  in order to ensure unidirectional transport and to avoid the influence of thermal fluctuations. The QPC detector was tuned to the edge of the first conduction step. The current through the QPC was measured with a sampling frequency of 500 kHz. The tunneling events were extracted from the QPC current using a step detection algorithm and converted into time-dependent pulse currents:  Time was discretized in steps of $\Delta t = 40$ $\mu$s and in each step the number of tunneling events $\Delta n$ in (out) of the QD was recorded. The current into (out of) the QD at a given time step is then $I(t)=(-e)\Delta n/\Delta t$.

\noindent {\bf Error estimates.} We estimated the errors of the measured spectra by dividing the experimental data into 30 separate batches. The spectra were determined for each batch individually and their standard deviation was used as a measure of the experimental accuracy.

\noindent {\bf Finite-frequency cumulants.} To define the finite-frequency cumulants of the current we consider the $m$-time probability distribution\cite{Emary2007,Marcos2010} $P^{(m)}(n_1,t_1;\ldots;n_m,t_m)$ that $n_k$ electrons have been transferred during the time span $[0,t_k]$ for $k=1,\ldots,m$. The corresponding cumulant generating function is
\begin{equation}
\mathcal{F}^{(m)}(\boldsymbol{\chi},\boldsymbol{t})\equiv\log\left\{\sum_\mathbf{n}P^{(m)}(\boldsymbol{n};\boldsymbol{t})e^{i \boldsymbol{n}\cdot\boldsymbol{\chi}}\right\}
\nonumber
\end{equation}
with $\boldsymbol{\chi}=(\chi_1,\ldots,\chi_m)$, $\boldsymbol{t}=(t_1,\ldots,t_m)$, and $\boldsymbol{n}=(n_1,\ldots,n_m)$. (An equivalent definition uses only a single but time-dependent counting field.\cite{Bagrets2003}) The $m$-time cumulants of $P^{(m)}(\boldsymbol{n};\boldsymbol{t})$ are then
\begin{equation}
\llangle n(t_1)\cdots n(t_m)\rrangle=\partial_{i\chi_1}\cdots\partial_{i\chi_m} \mathcal{F}^{(m)}(\boldsymbol{\chi},\boldsymbol{t})|_{\boldsymbol{\chi}\rightarrow \boldsymbol{0}},
\nonumber
\end{equation}
with the corresponding $m$-time cumulants of the current
\begin{equation}
\llangle I(t_1)\cdots I(t_m)\rrangle=\partial_{t_1}\cdots\partial_{t_m}\llangle n(t_1)\cdots n(t_m)\rrangle.
\nonumber
\end{equation}
These equations define the cumulant averages denoted by double brackets $\llangle\ldots\rrangle$. In the Fourier domain, the current cumulants can be expressed as
\begin{equation}
\llangle \hat{I}(\omega_1)\cdots \hat{I}(\omega_m)\rrangle=2\pi\delta(\omega_1+\ldots+\omega_{m})S^{(m)}(\omega_1,\ldots,\omega_{m-1})
\nonumber
\end{equation}
since the sum of frequencies is zero in the stationary state. The Fourier transformed current is denoted as $\hat{I}(\omega)$ and $S^{(m)}(\omega_1,\omega_2,\ldots,\omega_{m-1})$ is the polyspectrum,\cite{Nikias1993} which for $m=2$ and $m=3$ yields the noise spectrum $S^{(2)}(\omega)$ (second cumulant) and the bispectrum (third cumulant) $S^{(3)}(\omega_1,\omega_2)$ respectively. We note that the noise spectrum $S^{(2)}(\omega)$ in the stationary state is a single-frequency quantity which can be related directly to the one-time probability distribution $P^{(1)}(n,t)$ via MacDonald's formula\cite{MacDonald1949} $S^{(2)}(\omega)=\omega\int_0^\infty dt\sin{(\omega t)}\frac{d}{dt}\llangle n^2\rrangle (t)$.  The bispectrum $S^{(3)}(\omega_1,\omega_2)$ in contrast is a two-frequency quantity which reflects correlations of the current beyond what is captured by $P^{(1)}(n,t)$ alone.

\noindent{\bf Theoretical model.}
We consider the probability vector $|p(t)\rangle = [p_{00}(t),p_{10}(t),p_{01}(t),p_{11}(t)]^T$, where the first index denotes the number of (additional) electrons on the QD, $i=0,1$, and the second index denotes the detected number of (additional) electrons on the QD as inferred from the current through the QPC, $j=0,1$. The probability vector evolves according to the rate equation
\begin{equation}
\frac{d}{dt}|p(\chi_L,\chi_R;t)\rangle=\mathbf{M}(\chi_L,\chi_R)|p(\chi_L,\chi_R;t)\rangle,
\nonumber
\end{equation}
having introduced separate counting fields $\chi_L$ and $\chi_R$ that couple to the number of detected electrons that have passed the left and the right barriers, respectively.\cite{Bagrets2003} The matrix $\mathbf{M}(\chi_L,\chi_R)$ reads\cite{Naaman2006,Gustavsson2009,Flindt2009}
\begin{equation}
\left(
  \begin{array}{cccc}
    -\Gamma_L & \Gamma_R & \Gamma_De^{i\chi_R} & 0 \\
    \Gamma_L & -(\Gamma_R+\Gamma_D) & 0 & 0 \\
    0 & 0 & -(\Gamma_L+\Gamma_D) & \Gamma_R \\
    0 & \Gamma_De^{i\chi_L} & \Gamma_L & -\Gamma_R \\
  \end{array}
\right),
\nonumber
\end{equation}
where factors of $e^{i\chi_{L(R)}}$ in the off-diagonal elements correspond to processes that increase by one the number of detected electrons that have tunneled across the left (right) barrier, see Fig.\ \ref{Fig1}c,d. The detector rate $\Gamma_D$ of the QPC detector scheme tends to infinite for an ideal detector only, but is finite otherwise.

\noindent{\bf Calculations.} For the calculations of the finite-frequency cumulants it is useful to write the matrix as
\begin{equation}
\mathbf{M}(\chi_L,\chi_R)=\mathbf{M}_0+(e^{i\chi_L}-1)\mathbf{I}_L+(e^{i\chi_R}-1)\mathbf{I}_R,
\nonumber
\end{equation}
where $\mathbf{M}_0=\mathbf{M}(0,0)$ and $\mathbf{I}_{L(R)}$ is the super operator for the detected current through the left (right) barrier.\cite{Flindt2005} Additionally, we need the stationary state $|0\rrangle$, found by solving $\mathbf{M}_0|0\rrangle=0$ and normalized such that $\llangle\tilde{0}|0\rrangle=1$, where $\llangle \tilde{0}|=(1,1,1,1)$. We then define the projectors\cite{Flindt2005} $\mathbf{P}= |0\rrangle\!\llangle\tilde{0}|$ and $\mathbf{Q}=\mathbf{1}-\mathbf{P}$
and the frequency-dependent pseudoinverse\cite{Flindt2005b} $\mathbf{R}(\omega)= \mathbf{Q}[i\omega+\mathbf{M}_0]^{-1}\mathbf{Q}$,
which is well-defined even in the zero-frequency limit $\omega\rightarrow 0$, since the inversion is performed only in the subspace spanned by $\mathbf{Q}$, where $\mathbf{M}_0$ is regular. These objects constitute the essential building blocks for our calculations. The frequency-dependent second and third cumulants, $S^{(2)}(\omega)$ and $S^{(3)}(\omega_1,\omega_2)$, can then be evaluated following Refs.\ [\onlinecite{Flindt2005b,Flindt2008}] and [\onlinecite{Marcos2010}], respectively.
\newline

\noindent {\bf Acknowledgments}\\
\noindent We thank M.\ B\"{u}ttiker, C.\ Emary, and Yu.\ V.\ Nazarov for instructive discussions. W.\ Wegscheider (Regensburg, Germany) provided the wafer and B.\ Harke (Hannover, Germany) fabricated the device. The work was supported by BMBF via nanoQUIT (C.\ Fr., N.\ U., F.\ H., and R.\ J.\ H.), DFG via QUEST (C.\ Fr., N.\ U., F.\ H., and R.\ J.\ H.), the Villum Kann Rasmussen Foundation (C.\ Fl.), and the Swiss NSF (C.\ Fl.).
\newline

\noindent {\bf Author contributions}\\
\noindent All authors conceived the research. N.~U., C.~Fr., and F.~H.\ carried out the experiment and analyzed data. All authors discussed the results. C.~Fl.\ developed theory and performed calculations. N.~U., C.~Fr., and C.~Fl.\ wrote the manuscript. F.~H. and R.~J.~H. supervised the research. All authors contributed to the editing of the manuscript.
\newline

\noindent {\bf Additional information}
\newline
\noindent {\bf Corresponding author.} Correspondence and requests for materials should be addressed to R.\ J.\ H.\ (email: haug@nano.uni-hannover.de).

\end{document}